\documentstyle{mn}

\input{epsf}

\begin{document}

\title[Cosmological Perturbation Theory and the Spherical Collapse model]
{Cosmological Perturbation Theory and the Spherical Collapse model-
III. The Velocity divergence field \& the $\Omega$-dependence} 

\author[P. Fosalba and E. Gazta\~{n}aga]
{Pablo Fosalba and Enrique Gazta\~{n}aga \\ 
IEEC- Institut d'Estudis Espacials de Catalunya, Research Unit (CSIC), \\
 Edf. Nexus-201 - c/ Gran Capit\`a 2-4, 08034 Barcelona, Spain}

\maketitle
 
\def\Mpc{{\,h^{-1}\,{\rm Mpc}}}
\def\mpc {h^{-1} {\rm{Mpc}}}
\def\and  {\it {et al.} \rm}
\def\rmd {\rm d}

\def\etal {{\it et al., }}
\def\spose#1{\hbox to 0pt{#1\hss}}
\def\simlt{\mathrel{\spose{\lower 3pt\hbox{$\mathchar"218$}}
     \raise 2.0pt\hbox{$\mathchar"13C$}}}
\def\simgt{\mathrel{\spose{\lower 3pt\hbox{$\mathchar"218$}}
     \raise 2.0pt\hbox{$\mathchar"13E$}}}
\def\beq{\begin{equation}}
\def\eeq{\end{equation}}
\def\bce{\begin{center}}
\def\ece{\end{center}}
\def\bea{\begin{eqnarray}}
\def\eea{\end{eqnarray}}
\def\ben{\begin{enumerate}}
\def\een{\end{enumerate}}
\def\ul{\underline}
\def\ni{\noindent}
\def\nn{\nonumber}
\def\bs{\bigskip}
\def\ms{\medskip}
\def\wt{\widetilde}
\def\brr{\begin{array}}
\def\err{\end{array}}
\def\dsp{\displaystyle}
\newcommand{\rhobar}{\overline{\rho}}
\newcommand{\rhohat}{\hat{\rho}}
\newcommand{\xibar}{\overline{\xi}}
\newcommand{\deltabar}{\overline{\delta}}
\newcommand{\sigmabar}{\overline{\sigma}}
\newcommand{\deltahat}{\hat{\delta}}
\newcommand{\sigmahat}{\hat{\sigma}}
\def\Or{{\cal O}}
\def\calG{{\cal G}}
\def\La{{\cal L}}
\def\dotD {{\dot{D}}} 
\def\dotF {{\dot{F}}} 
\def\ddotF {{\ddot{F}}}
\def\ddotR {{\ddot{R}}}
\def\dotdel {{\ddot{\delta}}}
\def\ddotdel {{\ddot{\delta}}} 
\def\xdot {{\dot{\bf x }} }
\def\dota {{\dot{a}}} 
\def\vpsi {{\bf \Psi }}
\def\ie {{\it i.e.,}} 
\def\eg {{\it e.g.,}}

\font\twelveBF=cmmib10 scaled 1200
\newcommand{\bte}{\hbox{\twelveBF $\theta$}}
\newcommand{\x}{\hbox{\twelveBF x}}
\newcommand{\q}{\hbox{\twelveBF q}}
\newcommand{\vv}{\hbox{\twelveBF v}}
\newcommand{\y}{\hbox{\twelveBF y}}
\newcommand{\r}{\hbox{\twelveBF r}}
\newcommand{\k}{\hbox{\twelveBF k}}
\newcommand{\lexp}{\mathop{\bigl\langle}}
\newcommand{\rexp}{\mathop{\bigr\rangle}}
\newcommand{\rexpc}{\mathop{\bigr\rangle_c}{}}
\newcommand{\eq}{{equation~}}
\newcommand{\thetahat}{\hat{\theta}}

\begin{abstract}

Cosmological Perturbation Theory (PT) is a useful tool to 
study the cumulants of the density and velocity fields 
in  the large scale structure of the Universe.
In papers I \& II of this series 
we saw that the Spherical Collapse (SC) 
model provides with the exact solution to PT
at tree-level and gives a good approximation to the loop corrections
(next to leading orders), indicating negligible tidal effects.
Here, we derive predictions for the (smoothed) cumulants
of the velocity divergence field $\theta \equiv \,\nabla\cdot\vv$
for an irrotational fluid in the SC model.
By comparing with the exact analytic results 
by Scoccimarro \& Frieman (1996),
it is shown that, at least for the unsmoothed case, 
the loop corrections to the
cumulants of  $\theta$ are dominated by tidal effects. 
However, most of the tidal contribution 
seems to cancel out when computing the hierarchical ratios,
$T_J = {\langle \theta^J \rangle / \langle \theta^2 \rangle^{J-1}}$.
We also extend the work presented in Papers I \& II to give
predictions for the cumulants of the density and velocity divergence fields 
in non-flat spaces. In particular, we show the equivalence between the 
{\em spherically symmetric} solution to the equations of motion in the 
SC model (given in terms of the density) and that of the 
Lagrangian PT approach (in terms of the displacement field). 
It is shown that the $\Omega$-dependence is very weak for both cosmic
fields even at one-loop (a $10 \%$ effect at most), 
except for the overall factor $f(\Omega)$ that
couples to the velocity divergence.

\end{abstract}





\section{Introduction}

One of the aims of modern Cosmology is to understand
the origin and growth of the large scale structure 
as we observe it today.
Cosmological Perturbation Theory
(PT) has revealed to be a key analytic tool to make predictions about
the cosmological matter density $\delta$ and velocity $\vv$ fields in the 
weakly non-linear regime 
(\ie when $v, \delta \simlt 1$).
The first analytic results obtained were
only applicable to the {\em unsmoothed} fields, 
Gaussian initial conditions (GIC)
and flat space (\ie $\Omega = 1$ for a vanishing cosmological constant, 
$\Lambda $).
These results concern the tree-level (leading-order for GIC, see 
Peebles 1980, 
Fry 1984, Bernardeau 1992, hereafter B92).
Recently, loop corrections to the tree-level amplitudes 
(higher-order contributions for GIC) were obtained within the 
diagrammatic approach by
Scoccimarro \& Frieman (1996a, 1996b) although their results basically
concern the unsmoothed fields. 
Comparison with N-body simulations and observations  
made it necessary to incorporate the effects of
smoothing in the PT calculations 
(Goroff \etal 1986, Juszkiewicz \etal 1993, Bernardeau 1994a, 1994b,
Baugh, Gazta\~{n}aga,  Efstathiou 1995, Gazta\~{n}aga \& Baugh 1995).
Further results were obtained for non-Gaussian IC in a number of papers
(Fry \& Scherrer 1994, Chodorowsky \& Bouchet 1996)
which allows to discriminate
among models of structure formation (Silk \& Juszkiewicz 1991,
Gazta\~{n}aga \& M\"{a}h\"{o}nen 1996).

In Paper I (Fosalba \& Gazta\~naga 1998) and Paper II 
(Gazta\~naga \& Fosalba 1998)
of this series it was shown that {\em the Spherical Collapse (SC)
model gives the exact tree-level contribution to PT irrespective of the nature
of the IC and the geometry of the universe}. 
This means that, when it comes to computing the one-point cumulants at
tree-level, only the {\em spherically symmetric} solution to the equations
of motion is relevant (see Paper I \S 3.6). This solution 
generates the {\em monopole} term,
$\nu_n$, \ie the angle average of the kernels in Fourier space. 
As a result, the evolved density field is entirely determined by a local 
transformation of the linear density field alone, what we shall call a {\em 
local-density} transformation (see Paper I \S 3.4),
\beq
\delta  =  f(\delta_l) =  
\,\sum_{n=1}^{\infty} {\nu_n \over n!}\, [\delta_l]^n .
\label{loclag}
\eeq
Making use of this property we were able
to present in Papers I \& II results for Gaussian (GIC) and non-Gaussian IC 
(NGIC) respectively,
for both the unsmoothed and smoothed fields (for a top-hat window). 
Furthermore, the SC model was used to make predictions beyond the tree-level in
both cases which were shown to be in good agreement with N-body simulations 
up to the scales where PT itself is expected to break down (see
Paper I \S 5 and Paper II \S 3.3).

In this paper, we 
derive results for the peculiar velocity field $\vv$ which is the 
counterpart to the density field in the Newtonian equations of motion.
The peculiar velocity field as traced by a population of galaxies
has the potential advantage, with 
respect to the corresponding matter density, of being
a better (or at least different) tracer of the corresponding 
underlying field. This is because galaxy velocities should follow
the gravitational field regardless of their luminosity.
 These property has been proposed as a way
to make predictions about the density parameter
of the universe $\Omega$ (see Bernardeau 1994b)
and references therein).
In what follows we shall assume that the fluid is  
irrotational as is usually done in PT. This is a good
approximation in the SC model as the nonlinear density field is 
given in terms of the linear field alone, for which, PT predicts 
a dilution of vorticity with the expansion.
Within this approximation the velocity field is entirely determined
by its divergence,
$\theta \equiv \,\nabla\cdot\vv$. 
The cumulants 
(or connected moments) of the velocity divergence are defined as, 
${\langle \theta^J \rangle_c}$ 
and the corresponding hierarchical 
ratios are 
$T_J = {\langle \theta^J \rangle_c / \langle \theta^2 \rangle^{J-1}}$.
Results for the one-point cumulants of the 
velocity divergence field were given in Juszkiewicz \etal (1993) for
the skewness $T_3$ (at tree-level and GIC, alone). Bernardeau (1994a, 1994b,
B94a,b hereafter) systematically derived the
dependence of these statistical quantities on the initial power spectrum
for a top-hat window function
and the strong (overall)
omega dependence (see B94b) to leading-order (tree-level) for GIC.

We shall also extend the work to non-flat spaces (\ie  $\Omega \neq 1$ for a 
vanishing cosmological constant $\Lambda $) for the cumulants of
both the density and velocity
divergence fields. Curvature effects on the hierarchical amplitudes 
have been investigated in a number of papers: 
Martel $\&$ Freudling (1991),
and Bouchet \etal (1992) obtained the (weak) $\Omega$
dependence of the skewness of the density field in the 
Lagrangian PT approach.  Bernardeau (1994b, B94b hereafter) 
numerically integrated the $\Omega$-dependence
of the skewness and kurtosis of the density and velocity fields.
Exact analytic results , concerning the curvature-dependence
of the skewness, were derived by Bouchet \etal 1995 (BCHJ95 hereafter)
in Lagrangian PT and Catelan \etal 1995, in Euler space.
Recently, Nusser \& Colberg (1998) have investigated the
dependence on the cosmological parameters
$\Omega$ and $\Lambda$ in  the equations of motion and give some results
for the variance and the skewness.

This paper is organized as follows.
Section \S\ref{velfield} presents results 
for the velocity divergence 
field while 
in \S\ref{omega} results for  
the $\Omega$-dependence in the SC model are given. A 
discussion and
the final conclusions are given in \S\ref{discuss}.
Appendix \S\ref{ap:scomega} provides the connection between the
vertex generating function in PT and the SC equation for arbitrary $\Omega$. 
Appendix \S\ref{ap:omegadep} gives an alternative derivation of the $\Omega$
dependence in the monopole approximation to Lagrangian PT.

\section{Results for the Velocity Divergence Field}
\label{velfield}

In this section we apply our results from the SC model for the 
density field (see Papers I \& II) to derive 
the one-point cumulants for the
velocity field in the quasi-linear regime. 
For the current analysis, we shall consider an irrotational fluid and,
as a result, the velocity field is fully described by its divergence,
$\theta \equiv \nabla \vv /{\cal H}$, where ${\cal H} (\tau) = a(t) H(t)$
is the conformal Hubble parameter 
and $a(t)$ is the scale factor.
As mentioned above, this is a good approximation in the SC model as nonlinear
dynamics are fully determined by the linear growth of perturbations 
for which vorticity decays with time.

The velocity divergence is {\em locally} related to the linear 
{\it density} field
through the continuity equation, which for an Einstein-deSitter space reads, 
\beq
{\dot{\delta}}\,+\,(1+\delta)\,\theta\,= 0,
\label{conteq}
\eeq
where the dot denotes total time derivative. It follows that 
the linear solution is $\theta_l = - \delta_l$. 
This way, introducing Eq.[\ref{loclag}] above, we can obtain an analog
local-transformation for the velocity divergence, say,
\beq
\theta = g[\theta_l] = 
\sum_{k=1}^{\infty} {\mu_k \over k!}\, {[\theta_l]^k}
\label{locvel}
\eeq
where the {\em unsmoothed} $\mu_k$ coefficients 
are to be derived order by order
in a perturbative expansion of the continuity equation.  
Thus one finds,
\bea
\mu_1 &=& \nu_1 = 1 \nn \\
\mu_2 &=& 2\,(1-\nu_2) = -{26\over{21}} \simeq -1.24\nn \\
\mu_3 &=& 3\,(\nu_3\,-3\,\nu_2\,+2) = {142\over{63}} \simeq 2.25  \nn \\
\mu_4 &=& 4\,(-\nu_4\,+\,4\,\nu_3\,+3\,\nu_2^2\,-\,12\,\nu_2\,+\,6) \nn \\
&=& -{236872\over{43659}} \simeq -5.42 
\label{munus}
\eea
and so on.

\subsection{Smoothing Effects}

The smoothing effects for the velocity field for a top-hat window
can be easily derived by
applying similar arguments to those used for the density field (see Paper I
\S 4.4 and B94b).
In particular, the continuity equation ensures that the smoothed
velocity divergence field is proportional to the unsmoothed one, 
$\thetahat \sim \theta$ as induced by the sharp cut in the smoothed
density field with respect to the unsmoothed one, $\deltahat \sim 
\delta$. 
Moreover, the local character of the unsmoothed velocity
divergence field (see above) tells us that $\theta \sim g[\theta_l]$. Now,
from the linear continuity equation we get, $\thetahat_l 
= -\deltahat_l = - (\sigma_l / \sigmahat_l)\,\delta_l =
(\sigma_l / \sigmahat_l)\,\theta_l$. Introducing this last equality in the
local transformation $g$ which determines $\thetahat$ we finally get,
\beq
\thetahat \sim g[{\sigma_l \over \sigmahat_l}\,\thetahat_l]
\label{velsm}
\eeq
where $\sim$ just means that the transformation is performed in Lagrangian 
space so that it has to be properly normalized when going back to Euler space 
(see Paper I \S 4.2).

For a power-law power spectrum $P(k) \sim k^n$, one finds (see Paper I),
\beq
\sigmahat_l=\sigma_l \, 
(1+\deltahat)^{-\gamma/6}.
\eeq 
where $\gamma= -(n+3)$. As a result, the smoothed non-linear velocity
divergence field is given by,
\beq
\thetahat[\thetahat_l] \sim \theta[\thetahat_l  (1+\deltahat)^{\gamma/6}]
\label{thetahat} ,
\eeq  
to be normalized when transforming back to Euler space in an analogous way 
to the density field (see Paper I, Section 4.4). 
Notice that these results are 
formally equivalent to those provided by B94a for the smoothed fields in 
Euler space although, 
here, we work in Lagrangian space.  
 

According to this transformation, the first smoothed
coefficients of the velocity field read,
\bea
\overline{\mu_2} &=& {1\over 3}\,(-6\,\nu_2+6-\gamma) = \mu_2-
{\gamma\over 3} \nn \\
\overline{\mu_3} &=& 
{1\over 4}\,\left(4\,\mu_3-5\,\mu_2\,\gamma+\gamma^2 \right) \nn \\ 
\overline{\mu_4} 
&=& \mu_4\,- {20\over 9}\,\gamma\,\mu_3\,-\gamma\,\mu_2^2\, 
+ 2\,\gamma^2\,\mu_2 - {8\over 27}\,\gamma^3 , 
\label{musm}
\eea  
and so forth, where in the above equalities we have applied the 
identities given by 
Eq.[\ref{munus}]. 

Combining Eqs.[\ref{locvel}],[\ref{munus}] and [\ref{musm}], we can
derive the smoothed cumulants for the velocity divergence up to an arbitrary 
perturbative order in an Einstein-deSitter universe
for a top-hat window and a power-law power spectrum, for
GIC \& NGIC.

\subsection{Gaussian Initial Conditions}

Following the notation introduced in Paper I \S 5, we shall
use for the perturbative expansion of the variance,
\beq
\sigma_{\theta}^2 = \lexp \thetahat^2 \rexp = 
\sum_{i} {s_{2,i}^{\theta}\,\sigma_l^i}
\label{notat1}
\eeq
where $s_{2,1} \equiv 1$ and the subscript in the coefficients 
labels the order of the
perturbative expansion. Note that $\sigma_l$ denotes the
$rms$ fluctuation of the linear {\it density} field. 
As for the hierarchical 
amplitudes, we keep the above 
notation with the added labeling of the order of the moment,
$J$, that defines the $T_J$ coefficients, 
\beq
T_J \equiv
{\lexp \thetahat^J \rexp_c \over \lexp \thetahat^2 \rexp^{J-1}}\,= 
\sum_{i} {T_{J,i}\,\sigma_l^i}.
\label{notat2}
\eeq
As we did with the density field (see Paper I \S 5), we shall denote 
the leading order contributions (\ie the tree-level
for GIC) by $T_J^{(0)}$.

\subsubsection{Spherical Collapse Results}
\label{sec:scvel}

\begin{table}

\begin{center}

\begin{tabular}{|c||c|c|c|c|}
\hline \hline
SC & Unsmoothed & \multicolumn{3}{c|}{Smoothed} \\ 
\hline \hline
& $\gamma=0$ & $\gamma=-1$ & $\gamma=-2$ & $\gamma=-3$ \\ \hline
& $n=-3$ & $n=-2$ & $n=-1$ & $n=0$  \\ 
\hline \hline
$s_{2,4}^{\theta}$   & 10.97 & 7.98 & 5.61 &  3.84 \\ 
\hline \hline
$-T_{3,0}$ & 3.71 & 2.71 & 1.71 & 0.71 \\ 
\hline
$-T_{3,2}$ & 3.86 & -0.20 & -0.37 & 1.12 \\ 
\hline \hline
$T_{4,0}$ & 27.41 & 13.65 & 4.55 & 0.12  \\ 
\hline
$T_{4,2}$ & 131.06 & 29.97 & 17.49 & 15.30 \\ 
\hline \hline 

\end{tabular}

\caption[junk]{Values for the higher-order perturbative contributions in the
SC model for the unsmoothed ($n=-3$) and 
smoothed ($n=-2,-1,0$) velocity fields.}
\label{sc_vsm}
\end{center}

\end{table}

For GIC the odd terms in the perturbative
expansion vanish, thus,
\bea
\sigma_{\theta}^2 &=& \sigma_l^2 \,+\,s_{2,4}^{\theta}\,\sigma_l^4\,+
\,s_{2,6}^{\theta}\,\sigma_l^6\,+ \cdots  \nn \\
T_J &=& T_{J,0} \,+\,T_{J,2}\,
\sigma_l^2\,+\,T_{J,4}\,\sigma_l^4\,+ \cdots
\eea
For the variance, skewness and kurtosis of $\thetahat$
[for a top-hat window and a scale-free power spectrum, where
$n=-(\gamma+3)$] in an Einstein-deSitter universe, we find: 
\bea
s_{2,4}^{\theta} &=& {1613\over 147}\,+\, 
{415\over 126}\,\gamma \,+\, {11\over 36}\,\gamma^2 
\nn \\
-T_{3,0} &\equiv& -T_3^{(0)} = {26\over 7} \,+ \,\gamma \nn \\
-T_{3,2} &=& 
     {{43712}\over {11319}}\, +\, {{8926}\over {1323}}\,\gamma + 
        {55\over {18}}\,\gamma^2\,+\,{10\over {27}}\,\gamma^3 
\nn \\
T_{4,0} &\equiv& T_4^{(0)} = {{12088}\over {441}} \,
+\, {{338}\over {21}}\,\gamma \,+\, 
     {{7}\over 3}\,\gamma^2  \nn \\
T_{4,2} &=& {{10934570594}\over {83432349}}\,+\, 
        {{56796896}\over {305613}}\,\gamma \,+\, 
{{2613829}\over {23814}}\,\gamma^2 \nn \\
&+& {{31067}\over {1134}}\,\gamma^3 \,+\, {{1549}\over {648}}\,\gamma^4 .
\eea
The results for $T_{J,0}$ above are identical to those 
systematically obtained
by Bernardeau (1994a, 1994b) who made use of the 
{\em vertex generating function} formalism, although here they are
restricted to a power-law 
power spectrum. They can be straightforwardly extended to an arbitrary
(more realistic) power-spectrum by applying analog expressions to those
for the density field provided in Appendix A1 of Paper I.
They also extend those results to the one-loop contribution within
the SC approximation. 
Table \ref{sc_vsm} displays the results for the smoothed field for different 
values of the spectral index. 

To compare our results to those available 
in the literature we
should focus on the {\em unsmoothed} one-point cumulants,
\bea
\sigma_{\theta}^2 \approx \sigma_l^2\,+\,10.97\,\sigma_l^4
\,+\,{\cal O} (\sigma_l^6) \nn \\
-T_3 \approx 3.71\,+\,3.86\,\sigma_l^2\,+\,{\cal O} (\sigma_l^4) \nn \\
T_4 \approx 27.41\,+\,131.06\,\sigma_l^2\,+\,{\cal O} (\sigma_l^4).
\label{scvel}
\eea
The results from the exact PT in the diagrammatic approach 
derived by Scoccimarro \& Frieman (1996a) are,
\beq
\sigma_{\theta}^2 \approx \sigma_l^2\,+\,1.08\,\sigma_l^4
\,+\,{\cal O} (\sigma_l^6)
\label{ptvel}
\eeq
and the first corrective term for $T_3$ was viewed to strongly depend
on the spectral index due to non-local effects: $-2.58 \simgt T_{3,2}
\simgt -4.64$, for $2 \geq n \geq -2$ . Thus, our local value $T_{3,2} 
= -3.86$, is 
compatible with the average of the above non-local values. 
For the variance however, 
the corrective term seems to be dominated by non-local (tidal) effects 
which cut down non-linearities in an order of magnitude, unlike
the case of the density field (see Paper I \S 5.1). 
The situation might be different for
the smoothed cumulants, as we have seen (in Papers I\&II) that 
for $n \simeq -1.5$ tidal effects seem to cancel out.

\begin{figure}[t]
\centering
\centerline{\epsfysize=8.truecm 
\epsfbox{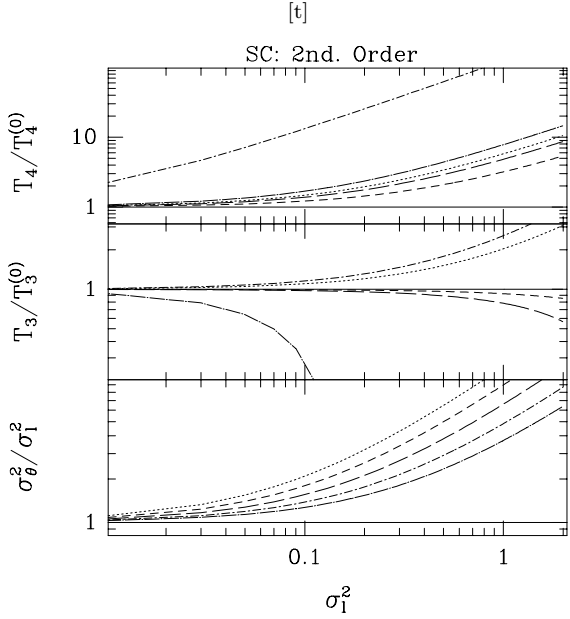}}
\caption[junk]{
Departures from the tree-level up to one loop
as a function of the linear rms fluctuation. 
The variance, skewness and kurtosis of the velocity divergence 
predicted by the SC model are shown for
different values of the spectral index: the dotted line shows the $n=-3$
(unsmoothed) case, and the short-dashed ($n=-2$), long-dashed ($n=-1$),
dot short-dashed ($n=0$), dot long-dashed ($n=1$) depict the 
behavior for the smoothed field. 
The solid line shows the tree-level values.}
\label{scvptv1}
\end{figure}

Fig \ref{scvptv1} depicts the departures from the tree-level contributions
for different values of the spectral index as the linear variance
grows. Notice that, in general, the smoothing effects tend to suppress
nonlinear contributions in accordance with the trend observed for
the density field (see Paper I Table 2 \& Paper II Table 1).           

\subsubsection{Spherically Symmetric Zel'dovich Approximation Results} 
\label{sec:SSZA}

\begin{table}

\begin{center}

\begin{tabular}{|c||c|c|c|c|}
\hline \hline
SSZA & Unsmoothed & \multicolumn{3}{c|}{Smoothed} \\ 
\hline \hline
& $\gamma=0$ & $\gamma=-1$ & $\gamma=-2$ & $\gamma=-3$ \\ \hline
& $n=-3$ & $n=-2$ & $n=-1$ & $n=0$  \\ 
\hline \hline
$s_{2,4}^{\theta}$   & 6.56 & 4.47 & 3 &  2.14 \\ 
\hline \hline
$-T_{3,0}$ & 2 & 1 & 0 & -1 \\ 
\hline
$-T_{3,2}$ & 2.07 & 1.43 & 2 & 1.57 \\ 
\hline \hline
$T_{4,0}$ & 8 & 1.67 & 0 & 3  \\ 
\hline
$T_{4,2}$ & 38.44 & 17.40 & 6 & -0.78 \\ 
\hline \hline 

\end{tabular}

\caption[junk]{Values for the higher-order perturbative contributions in the
SSZA for the unsmoothed ($n=-3$) and 
smoothed ($n=-2,-1,0$) velocity fields.}
\label{ssza_vsm}
\end{center}

\end{table}

The Zel'dovich Approximation (ZA hereafter, see Zel'dovich 1970)
formally corresponds to the first order in the Lagrangian approach to
PT (see Appendix \ref{ap:omegadep} for details) which is based
in a perturbative expansion of the Jacobian that relates the Lagrangian
to the Eulerian coordinates. It assumes that the particle positions in comoving
coordinates follow straight trajectories in Lagrangian space. 
In this section we proceed and give the results for the
{\em spherically symmetric} solution to the equations of motion
within the Zel'dovich Approximation 
(SSZA hereafter, see Paper I Appendix A4). In particular, we give estimates 
of the cumulants for the {\em smoothed} velocity divergence (for a top-hat
window and a power-law power spectrum)
in an Einstein-deSitter universe,

\bea
s_{2,4}^{\theta} &=& {59\over 9}\,+\, 
{43\over 18}\,\gamma \,+\, {11\over 36}\,\gamma^2 
\nn \\
-T_{3,0} &\equiv& -T_3^{(0)} = 2 \,+ \,\gamma \nn \\ 
-T_{3,2} &=&  {{56}\over {27}}\, +\, {2}\,\gamma + 
        {31\over {18}}\,\gamma^2\,+\,{10\over {27}}\,\gamma^3 
\nn \\
T_{4,0} &\equiv& T_4^{(0)} = 8 \,+\, {{26}\over {3}}\,\gamma \,+\, 
     {{7}\over 3}\,\gamma^2 \, \nn \\
T_{4,2} &=& {{346}\over {9}}\,+\, 
        {{3392}\over {81}}\,\gamma \,+\, 
{{5447}\over {162}}\,\gamma^2 \nn \\
&+&   {{2459}\over {162}}\,\gamma^3 \,+\, {{1549}\over {648}}\,\gamma^4 
\eea
thus, the {\em unsmoothed} cumulants ($\gamma=0$) show the following scaling,
\bea
\sigma_{\theta}^2 \approx \sigma_l^2\,+\,6.56\,\sigma_l^4
\,+\,{\cal O} (\sigma_l^6) \nn \\
-T_3 \approx 2\,+\,2.07\,\sigma_l^2\,+\,{\cal O} (\sigma_l^4) \nn \\
T_4 \approx  8\,+\,38.44\,\sigma_l^2\,+\,{\cal O} (\sigma_l^4).
\eea
This should 
be compared to the {\em unsmoothed} results from the exact PT 
in the diagrammatic approach 
up to the one-loop contribution (see Scoccimarro \& Frieman 1996a),
\bea
\sigma_{\theta}^2 \approx \sigma_l^2\,+\,0.73\,\sigma_l^4
\,+\,{\cal O} (\sigma_l^6) \nn \\
-T_3 \approx 2\,+\,1.64\,\sigma_l^2\,+\,{\cal O} (\sigma_l^4) \nn \\
T_4 \approx 8\,+\,17.58\,\sigma_l^2\,+\,{\cal O} (\sigma_l^4)
\eea
illustrating again the point made above (see \S\ref{sec:scvel}): 
tidal effects dominate
the variance of the velocity field and make non-linear effects small,
what is at variance with predictions from the local
SSZA dynamics. 
Again here, the situation might be different for
the smoothed cumulants, as we have seen (in Papers I\&II) that 
for $n \simeq -1.5$ tidal effects seem to cancel out.
Despite this limitation 
intrinsic to our {\em local} approximation, we see that tidal effects 
almost cancel their
contributions to the reduced cumulants.
This cancellation renders the SSZA approximation 
a good estimator of these quantities within the ZA dynamics,
in line with  the results obtained previously
for the density field. Of course the results in Eq.[\ref{scvel}] are
a better approximation to the {\it exact} PT.

\begin{figure}[t]
\centering
\centerline{\epsfysize=8.truecm 
\epsfbox{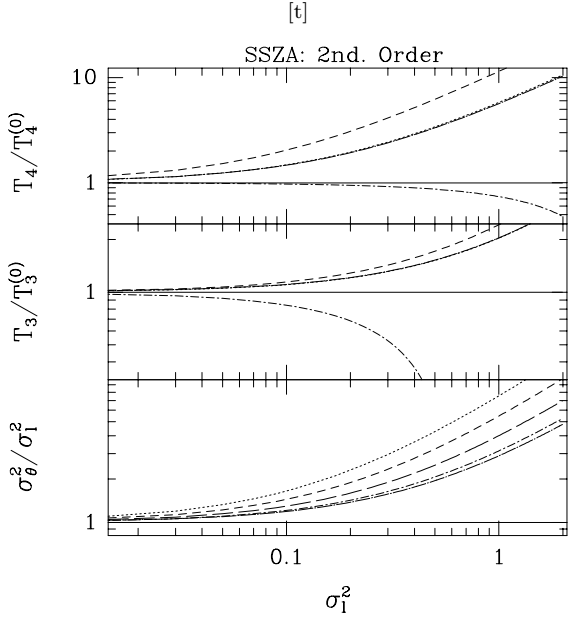}}
\caption[junk]{Same as Fig \ref{scvptv1} for the SSZA. Note that
the line $n=-1$ is not shown for $T_3$ and $T_4$ since the tree-level 
hierarchical amplitudes vanish. The values $n=-3$ (unsmoothed)
and $n=-1$ lead to the same scaling properties for the skewness and 
kurtosis so the corresponding lines overlap.}
\label{scvptzav1}
\end{figure}

Table \ref{ssza_vsm} shows the results for the smoothed velocity divergence
field in the SSZA for different values of spectral index.
Fig \ref{scvptzav1} displays deviations from the tree-level contributions
for different values of $n$ as the linear variance approaches unity.

\subsection{Non-Gaussian Initial Conditions}


For NGIC, no additional equations need to be solved as all the information
relevant for the computation of the one-point cumulants is encoded 
in the Gaussian tree-level (see Paper I \S 3.8).   
The perturbation 
expansion is worked out in an analog way to that for GIC, the only 
difference being the order at which the dominant terms appear, as induced
by the scaling properties of the IC (see Paper II \S 4).  
We represent the different orders in the expansion
as we did for the GIC above, Eq.[\ref{notat1}]-[\ref{notat2}].

We will assume, as a generic case, that the NGIC follow 
a {\em dimensional} scaling: ${\langle \theta_l^J \rangle}_{c} 
\propto {\langle \theta_l^2 \rangle}^{J/2}$, and we will
define the corresponding initial amplitudes as:
\beq
B_J \equiv B_J^{\theta} = {\lexp \thetahat_l^J \rexp_c 
\over{ {\langle \thetahat_l^2 \rangle}^{J/2}}}, 
\eeq
not to be confused here with the corresponding coefficients 
for the  density field.
This scaling induces different ordering in the perturbative 
series as compared to the GIC one. In particular,  
we find the following results for the first contributions to the
cumulants filtered with a top-hat
(assuming a power-law spectrum) and an Einstein-deSitter background,
 
\bea
s^{\theta}_{2,3} &=& -\,B_3\,\left({T_3^G\over 3}+1 \right)
\nn \\
-T_{3,-1} &\equiv& -T_3^{(0)} = B_3 \nn \\
-T_{3,0} &=& -T_3^G + 2\,\left({T_3^G\over 3}+1\right)
\,B_3^2 -  
\left({T_3^G\over 2}\,+1\right)\,B_4
\nn \\
T_{4,-2} &\equiv& T_4^{(0)} = B_4 \nn \\
T_{4,-1} &=& -4\,T_3^G\,B_3 + 
\left(3 + T_3^G\right)\,B_3\,B_4 \nn \\
&-& \left({2\over 3}\,T_3^G + 1\right)\,B_5 ,
\eea
where $T_J^G$ correspond to the {\em tree-level} PT results ($T_{J,0}$)
for GIC.
The {\em unsmoothed} one-point cumulants (setting $\gamma=0$
in the Gaussian coefficients above) we find:
\bea
s_{2,3}^{\theta} &=& {5\over 21}\,B_3\,\sigma_l^3 \nn \\
-T_{3,-1} &\equiv& -T_3^{(0)} = B_3 \nn \\
-T_{3,0} &=& {26\over 7}\,-\,{10\over 21}\,B_3^2\,+\,
{6\over 7}\,B_4 \nn \\
T_{4,-2} &\equiv& T_4^{(0)} = B_4 \nn \\
T_{4,-1} &=&
-{104\over 7}\,B_3\,-\,{5\over 7}\,B_3\,B_5
\,+\,{31\over 21}\,B_5 
\label{t_jdim}
\eea
We can compare these results to those from exact PT, derived
by Protogeros and Scherrer (1997), 
which give for $T_3$ for the unsmoothed
fields (setting the overall $\Omega$-dependence, $f(\Omega) =1$ for a
flat universe),
\bea
-T_3 &=& {\langle \delta_l^3 \rangle \over \sigma_l^4}\,
+\,{26\over 7}\,-\,{10\over 21}\,{\langle \delta_l^3 \rangle^2 
\over \sigma_l^6}\,+\,
{6\over 7}\,{\langle \delta_l^4 \rangle \over \sigma_l^4} \nn \\
&+&\,{4\over 7}\left[3\,{I[\xi_4^0]\over \sigma_l^4}\,  
-\,4\,{\langle \delta_l^3 \rangle\,I[\xi_3^0] \over \sigma_l^6}\right] 
\,+\,{\cal O} \left( \sigma_l \right) ,
\eea
being $I[\xi_J^l]= D(t)^J I[\xi_J^0]$, where $D(t)$ is the linear
growth factor and the
$I[\xi_J^0]$ denote some integral of the initial
$J$-point functions which are intrinsically {\em non-local} terms 
(see Eq.[30] in Paper II).
Thus, using $T_3^G=-26/7$ and $B_J = 
{\langle \theta_l^J \rangle_c}/\sigma_l^J$ in our expressions (see
Eq.[\ref{t_jdim}] above), we find, as we did
with the density field (see Papers I \& II), that the SC model 
gives the exact result up to some 
tidal (non-local) terms which result in 
a small contribution to the hierarchical amplitudes. This
smallness is partially due to the fact that they
enter as a difference between integrals of the initial $J$-point functions
which are generically of the same order, at least for the density field
(see Paper II, Section 4.4).

 For the smoothed fields however, our results must be taken as a prediction 
since there are no explicit results for non-Gaussian initial conditions. 
For some general implicit results, which depend on the
initial $J$-point functions for the smoothed $T_3$, see Protogeros
\& Scherrer (1997). This
should contain all our local terms when given in a somewhat more explicit
fashion.

\begin{figure}[t]
\centering
\centerline{\epsfysize=8.truecm 
\epsfbox{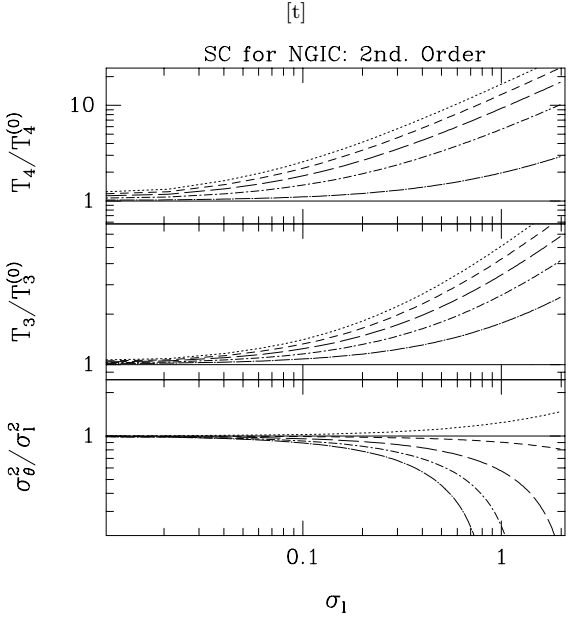}}
\caption[junk]{Same as Fig \ref{scvptv1} for non-Gaussian dimensional
initial conditions. We assume
$B_J \approx 1$.}
\label{scvpttopv1}
\end{figure}

\begin{figure}[t]
\centering
\centerline{\epsfysize=8.truecm 
\epsfbox{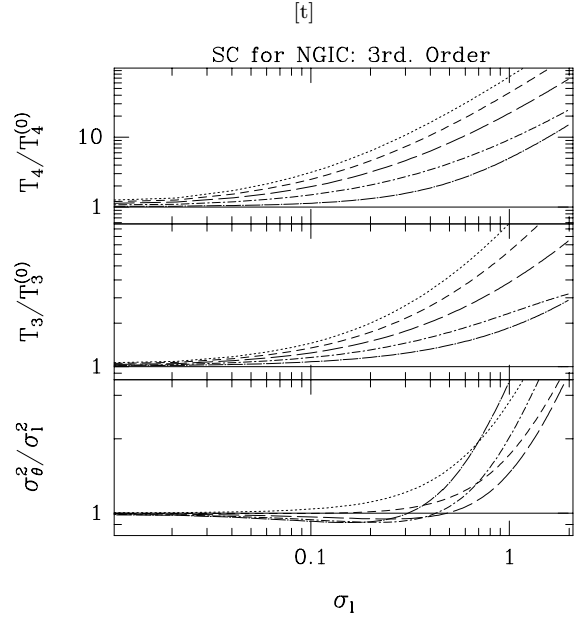}}
\caption[junk]{Same as Fig \ref{scvpttopv1} when the 3rd. 
perturbative contributions
are taken into account in the evolved one-point cumulants.} 
\label{scvpttopv2}
\end{figure}

Figs \ref{scvpttopv1} and \ref{scvpttopv2} show the evolved variance, 
skewness and kurtosis
as they depart from their first perturbative contributions $T_J^{(0)}$. 
As we pointed out in the case of the density
field (see Paper II, Section 3.1, Figs. 1 \& 2), the variance of 
the velocity divergence is seen to be dominated 
by the initial conditions in
the second perturbative contribution (only terms $\sim B_J$ appear),
while gravity takes over once the 
third perturbative contribution is included. This is realized by
the enhancement of the non-linear variance with respect of the linear one
(at least, for large values of the linear $rms$ density fluctuation)
in accordance with the expectation from the SC model for GIC.  
The hierarchical amplitudes
show the same coupling to the initial conditions as for the 
density field so all perturbative contributions are taken over by the 
initial conditions.
In Figures \ref{scvpttopv1} \& \ref{scvpttopv2} we have taken the initial 
{\em dimensional} coefficients $B_J \approx 1$, 
in analogy to the case for the
density field (see Paper II \S 3.1), although we find no physical motivation 
for why they should be 
comparable. However, we have checked that the qualitative behavior of the
statistical quantities and thus the overall picture is not significantly 
altered if we change those amplitudes by an order of magnitude \ie our
conclusions are robust. 
 
\section{The $\Omega$-dependence in the Spherical Collapse Model}
\label{omega}

\subsection{The $\Omega$-dependence in the Density Field}

\subsubsection{The Spherical Collapse Model Approach}
\label{omscapp}
 In the previous papers (see Papers I \& II) we  assumed a 
spatially flat (Einstein-deSitter) universe with a 
vanishing cosmological constant $\Lambda$, in order to
derive our predictions for the one-point cumulants. Although it is the
simplest analytic approach and there are strong theoretical arguments
which favor an $\Omega=1$, $\Lambda=0$ universe 
(\eg standard inflationary models),
it remains to be seen  whether or not this is actually so. 

In our calculations, we have neglected the effect of a non-vanishing
cosmological constant $\Lambda$. However, 
detailed analyses have shown   
that the dependence of the cumulants on this parameter
is extremely weak anyway, at least for a flat universe $\Omega + \Lambda =1$
(see Lahav \etal 1991, BCHJ95).

We showed in Paper I that in the general case of a perturbative approach
to estimate the leading order (tree-level) contribution to the cumulants,
it is  not necessary to calculate the full 
kernels $F_n$, but only the 
{\it monopole} contribution, $\nu_n \equiv <F_n>$, which corresponds
to the spherically symmetric (angle) average.
Thus the values of $\nu_n$  can be determined by finding
the spherically symmetric solution to the equations of motion.
We also show that for the gravitational evolution 
of the density field, this solution is given by 
the spherical collapse (SC). The proof (\S 4.1 in Paper I)
holds irrespective of the geometry of the universe, 
\ie for arbitrary $\Omega$ (which shows explicitly through the
Friedmann equation: $4 \pi G \bar{\rho} = 3/2 \Omega H^2$).
Therefore, we can write the local non-linear transformation introduced
in flat space (see Paper I, \S 4 Eq.[36]) and write, 
\beq
\delta = \sum_{i=1}^{\infty}\,{\nu_n (\Omega) \over n!}\,[\delta_l]^n ,
\eeq
where the  $\nu_n (\Omega)$ coefficients are a 
generalization of the flat-space monopole amplitudes, $\nu_n$, to non-flat
FRW geometries. This can be readily obtained by considering the
parametric solution to the SC dynamics in an 
open universe ($\Omega < 1$):

\bea
\delta_l &=& D(\psi)\,\left[ \left({\sinh \theta -\theta  \over
\sinh \psi-\psi} \right)^{2/3}\,-\,1 \right] \nn \\
\delta &=& \left({\cosh \psi-1 \over \cosh \theta -1}\right)^3 \times
\left({\sinh \theta -\theta  \over \sinh \psi-\psi} \right)^2\,-\,1  \nn 
\eea
for $\delta_l > -D(\psi)$, and
\bea
\delta_l &=&-D(\psi)\,\left[ \left({-\sin \theta +\theta  \over
\sinh \psi-\psi} \right)^{2/3}\,+\,1 \right] \nn \\
\delta &=& \left({\cos \psi-1 \over -\cos \theta +1}\right)^3 \times
\left({\sin \theta -\theta  \over \sinh \psi-\psi} \right)^2\,-\,1 ,
\eea
for $\delta_l < -D(\psi)$, with
\bea
D(\psi) &=&{9\over 2}\,{\sinh \psi (\sinh \psi-\psi) \over 
(\cosh \psi -1)^2}-3 \nn \\
\psi &=& \cosh^{-1} \left({2\over \Omega}-1 \right) ,
\eea
where $D(\psi)$ is the growing mode of the linear density contrast
(see also B92). 
In order to get the solution for an $\Omega >1$ universe, the transformation
$\psi \rightarrow i \psi$ must be performed in the above expressions.
This way, the $\nu_n (\Omega)$ coefficients may be numerically integrated
by expanding the parametric solutions around $\delta_l = 0$.
An alternative derivation of the $\Omega$-dependence of the Gaussian 
tree-level was presented by Bernardeau (1992) in terms of the vertex
generating function. In Appendix A1 we present the connection 
between the latter and the SC equation for arbitrary $\Omega$.

A good fit to the above results can be obtained
by realizing that in Lagrangian PT (see \S\ref{ap:omegadep})
the $\Omega$-dependence of the hierarchical amplitudes only appears beyond
the first perturbative contribution (which corresponds to the ZA). As the larger the
density parameter is, the smaller the $S_J$ parameters are (this is a 
monotonic behavior, at least within the range $10 \simgt \Omega \simgt 0$), 
one is led
to the simplest possible Ansatz for $S_J (\Omega)$, which is of the form,
\bea
S_J (\Omega) &=& S_J^{ZA} + B\, \Omega^{-\alpha} \nn 
\eea
being $\alpha$ and $B$ positive constants. 
Imposing that for $\Omega=1$ one must recover the well-known
Einstein-deSitter values, we  find:
\beq
S_J (\Omega) = S_J^{ZA} + \left(S_J^{\Omega=1}-S_J^{ZA} \right) 
\Omega^{-\alpha} 
\label{ansatz} 
\eeq
so that $\alpha$ is the only parameter to be fitted by inspection
in a log-log plot of the numerically integrated solutions. 
In particular, for the cosmologically favored range 
$2 \simgt \Omega \simgt 0.1$, 
we obtain the following numerical fits for
the tree-level hierarchical amplitudes,
\bea
S_3 (\Omega) &=& 3\,\nu_2 (\Omega) \approx {34\over 7} + 
{6\over 7}\,\left( \Omega^{-3/91}-1 \right) \nn \\
S_4 (\Omega) &=& 4\,\nu_3 (\Omega)+12\,{\nu_2 (\Omega)}^2 \nn \\
&\approx&
{60712\over 1323} + {20728\over 1323}\,\left( \Omega^{-1/27}-1 \right) 
\label{sjnuom}   
\eea
from which we derive the fits for $\nu_2 (\Omega)$ and $\nu_3 (\Omega)$. 
Replacing these unsmoothed coefficients in the general expressions
for the smoothed $\overline\nu_k$  
(since the smoothing effects were derived regardless
of the geometry of the universe, see Paper I \S 4.4), we get
at tree level,
\bea
S_3 (\Omega) &=& 3\,\overline{\nu_2(\Omega)} \approx {34\over 7} + 
{6\over 7}\,\left( \Omega^{-3/91}-1 \right) + \gamma \nn \\
S_4 (\Omega) &=& 4\,\overline{\nu_3 (\Omega)} + 12\,{\overline{\nu_2
(\Omega)}}^2 \nn \\
&\approx &  {60712 \over 1323} +{20728\over 1323}\,
\left(\Omega^{-1/27}-1 \right)+  \nn \\
&+& {62\over 3}\,\gamma+{7\over 3}\,\gamma^2  
+4\,\gamma\,\left(\Omega^{-3/91}-1 \right) .
\label{sjg}
\eea

\begin{figure}
\centering
\centerline{\epsfysize=8.truecm 
\epsfbox{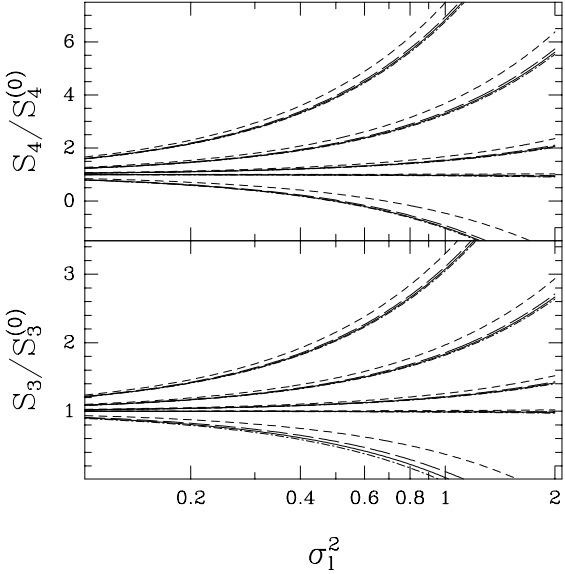}}
\caption[junk]{Predictions for the departures from the tree-level contributions
to the skewness and kurtosis for non-flat universes. The solid line corresponds
to the flat-space prediction while the short-dashed, long-dashed and dot-dashed
lines give the behavior for models with $\Omega = 0, 0.4, 2$ respectively.
It is seen how the weak dependence on
with the density parameter of the non-linearities, makes it hard to distinguish
them from those of flat-space unless $\Omega$ approaches zero. 
The upper group
of lines corresponds to $n=-3$, while lower groups of lines show the smoothed
cases: from  n=-2 to n=1 (bottom group of lines).}
\label{nflatscal}
\end{figure}

Moreover, the numerical fits for the $\nu_n (\Omega)$ can be used to derive
the corresponding approximate $\Omega$-dependences for the $\sigma$-corrections
to the cumulants as they carry all the dynamical information 
in the SC model (see
Paper I, \S 3.5). Fig \ref{nflatscal} shows the departures from the 
tree-level amplitudes
for different values of $\Omega$. As shown there, the dependence is 
rather weak and only for values of $\Omega$ close to zero are the predictions
a $10 \%$ greater than those for flat space.  

We have also explored the $\Omega$-dependence of the non-linear variance.
Making use of the fits derived for the $\nu_k (\Omega)$ above, we find for
the first non-linear correction (one-loop),
\bea
s_{2,4} &=& {1909\over 1323} + {143\over 126}\,\gamma +{11\over 36}\,\gamma^2
-{10\over 49}\,\left(\Omega^{-6/91}-1 \right) \nn \\
&\hspace{-0.8cm} +&\hspace{-0.8cm}
{5182\over 1323}\,
\left(\Omega^{-1/27}-1 \right) + {1\over 21}\,\left(11\,\gamma-64 \right)
\left(\Omega^{-3/91}-1 \right) .
\eea

From the above result we estimate that the non-flat ($\Omega <1$)
 variance to one-loop 
can be a $10 \%$ greater at most (for $\sigma_l<1$),
than the flat space value, similarly to
what we found for the $S_J$ amplitudes. 
We stress nevertheless
that the SC result for the variance only gives an accurate
approximation to the exact PT result (and the N-body predictions) around 
$n \simeq -1.5$ ($\gamma \simeq -1.5$), where tidal forces yield a 
negligible contribution (see Paper I \S 5.2).

The general trend observed in the hierarchical amplitudes for the 
density contrast in the non-linear
regime, according to the SC model, is that, the lower the density 
parameter is, the higher
the amplitudes are. 
This reflects the fact that non-linear terms in the perturbative expansion 
of the overdensity in the low $\Omega$ model are larger at a given time 
than the corresponding ones in flat space. 
This seems in qualitative agreement with the results recently 
found by Nusser \& Colberg (1998)
concerning the variance and the skewness from N-body simulations, for
standard CDM models (see their Table 1). A direct comparison is difficult 
to do in a significant way given that
there are no errors, which are important for this slight deviations,
and the quoted scales lie close to the nonlinear regime
$\sigma \simgt 1$.

\subsubsection{The Lagrangian PT Approach}

An independent derivation of the approximated $\Omega$-dependence in the 
hierarchical amplitudes of the density and velocity fields is provided by 
the Lagrangian PT approach (see Moutarde \etal 1991, BCHJ95). 
Since the {\em spherically symmetric} solution
to the equations of motion gives
the exact tree-level contribution to PT (see Paper I \S 3.4 \& 
Paper II \S 2.4), 
we shall impose the 
{\em spherically symmetric} condition on the 
relevant equations of motion in Lagrangian PT to get the exact tree-level 
as well in that formalism. Lagrangian PT is formulated 
in terms of the perturbation of the original trajectory in Lagrangian
space $q$, so that the perturbative series is built 
by expanding the Jacobian $J=| {\partial x / \partial q}|$ 
of the transformation between Lagrangian and Euler
coordinates, $x(t,q) = q + \Psi (t,q)$ (being $\Psi$ the 
so-called displacement field). This is equivalent to expanding
the density contrast in Lagrangian space since,
\beq
\delta = {\rho \over \overline{\rho}}-1 = {1\over J}-1 .
\eeq
 
In this framework, the derivatives of the
displacement field, $\Psi_{i,j}$, 
play the role of the smallness parameter in the perturbative expansion,
instead of the linear density contrast, $\delta_l$, as 
it was the case for the SC model.
By imposing {\em spherical symmetry} in the growth of perturbations we
set $\Psi_{1,1} =  \Psi_{2,2} = \Psi_{3,3} = \Psi_{k,k}$ 
(being zero otherwise). So 
for all perturbative orders, it holds that, $J^{(n)} \propto 
[J^{(1)}]^{n}$ (see \S\ref{ap:omegadep} for details). This is the key point
in the connection with the SC model in the perturbative regime 
as it explicitly shows that all the
perturbative orders can be expressed as powers of the linear one as was the 
case for the expansion in terms of the linear density field in the SC model.
In other words, we can build a local-density transformation 
(see Eq.[\ref{loclag}])
in Lagrangian
PT from the {\em spherically symmetric} solution to the equations of motion
whose coefficients (the $g_k$ functions, see \S\ref{ap:omegadep}) 
carry the same 
information than the $\nu_k$ ones in the
SC model. Identifying the linear terms in both expansions we can
formally relate all higher perturbative orders one by one, from which we
can derive the relations between the coefficients ($g_k$ with $\nu_k$)
that determine the one-point
statistics in both formalisms. 

In particular (See Appendix A2), we get for the first coefficients, 
\bea
\nu_2 &=& {2 \over 3}\,\left(2 - {g_2 \over g^2_1} \right) \nn \\
\nu_3 &=& {2 \over 9}\,\left(10 - 12\,{g_2 \over g^2_1} + 
{g_{3a} \over g^3_1} + 6\,{g_{3b} \over g^3_1} \right).
\label{nus}
\eea
Once we relate
both {\em unsmoothed} coefficients $\nu_k$ with $g_k$, we can simply
apply all previously given results for the SC model by rewriting
the expressions appropriately in terms of the new $g_k$ coefficients.
It is important to stress the equivalence between the SC model predictions in 
Lagrangian and Euler space at tree-level (see Paper I \S 4.2). Thus
only  beyond tree-level (next-to-leading terms in PT) there are differences 
between the SC model formulated in Euler space and
the {\em spherically symmetric} approach to Lagrangian PT.

Furthermore, the equivalence between both formalisms (the SC model
in the perturbative regime and the Lagrangian PT) 
in Lagrangian space, allows to induce a 
fit to the $\Omega$-dependence of the
$g_k$ functions 
from that of the $\nu_k$ ones. 
We can compare the fits  given by BCHJ95 to
our fits for the cumulants to see which ones match better
the numerical results.
Although the exact analytic expressions 
for $g_1$ \& $g_2$ are known (see BCHJ95),
no exact results are available for higher-orders, so we
shall focus on their  numerical fits for the cosmologically favored 
parameter space for $\Omega$ and work out the predictions. According to
BCHJ95, near $\Omega = 1$, good fits are provided by the following factors,
\beq
{g_2 \over g_1^2} \approx  -{3\over 7}\,\Omega^{-2/63}, \quad
{g_{3a} \over g_1^3} \approx  -{1\over 3}\,\Omega^{-4/77}, \quad
{g_{3b} \over g_1^3} \approx  {5\over 21}\,\Omega^{-2/35} .
\label{gs}
\eeq
Combining Eqs.[\ref{nus}] \& [\ref{gs}] we get the  
fits to the $\nu_k (\Omega)$ coefficients from 
the Lagrangian PT approach. 
The smoothing effects for a top-hat window can be incorporated as
described in Paper I, Section 4.4. The resulting smoothed coefficients
$\overline{\nu_k (\Omega)}$ in terms of the unsmoothed ones 
$\nu_k (\Omega)$, are given by Eq.[A3] in Paper I.
Replacing them in the smoothed $S_J$ amplitudes yields, for a power-law 
spectrum [with $\gamma = -(n+3)$],
\bea
S_3 (\Omega) &=& 3\,\overline{\nu_2} \approx {34 \over 7} + {6\over 7}
\,\left( \Omega^{-2/63}-1 \right) + \gamma \nn \\
S_4 (\Omega) &=& 4\,\overline{\nu_3} + 12\,\overline{\nu_2}^2 \nn \\ 
&\approx & {60712 \over 1323} +{48\over 49}\,\left(\Omega^{-4/63}-1 \right)
+ {80\over 63}\,\left(\Omega^{-2/35}-1 \right) \nn \\  
&-& {8\over 27}\,
\left(\Omega^{-4/77}-1 \right) \nn \\
&+& {62\over 3}\,\gamma+{7\over 3}\,\gamma^2
 + \left({96\over 7}+4\,\gamma \right)
\left(\Omega^{-2/63}-1 \right) .
\label{sjlpt}
\eea



\begin{figure}
\centering
\centerline{\epsfysize=8.truecm 
\epsfbox{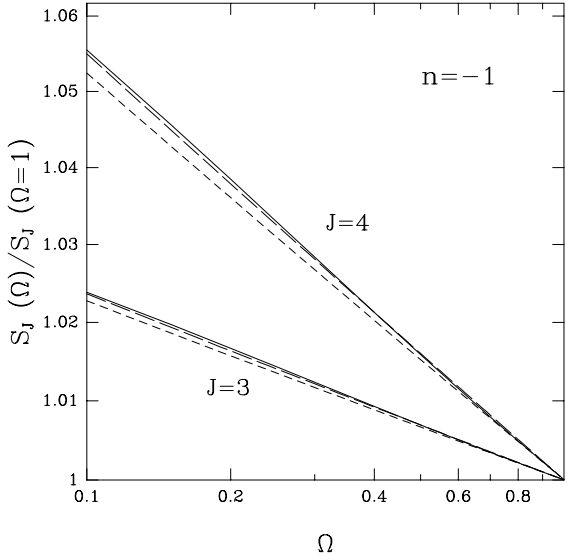}}
\caption[junk]{
Comparison between the exact (numerically integrated)
$S_J (\Omega)$ amplitudes (normalized to the flat space values) at tree-level 
(solid line) and the numerical fits
to the $\Omega$-dependence for open models (dashed lines). The 
case $n=-1$ is displayed. The 
short-dashed line shows the SC predictions making use of the 
BCHJ95 fits to the $g_k$ functions, Eq.[\ref{gs}], in the 
context of the Lagrangian PT, while the long-dashed line gives the
fits to the SC model solutions according to the Ansatz Eq.[\ref{ansatz}].} 
\label{nftree_den}
\end{figure}

Fig \ref{nftree_den} shows the $\Omega$-dependence 
of the skewness and kurtosis of the density field at tree-level, which
is shown to be very weak, as expected. In particular we compare 
the exact solution for the $\Omega$-dependence to the fits to the 
exact behavior 
according to the SC model approach (based on Eq.[\ref{ansatz}]) on one hand, 
and those based on the 
{\em spherically symmetric} solution to the Lagrangian PT
(for which we make use of the BCHJ95 fits to the $g_k$ functions, 
Eq.[\ref{gs}]).
Our fits are slightly more accurate 
than those given by BCHJ95, 
which nevertheless only underestimate the exact $\Omega$-dependence
of the skewness, $S_3$, and kurtosis, $S_4$, by
$3 \%$ and $5 \%$ at most.

\subsection{The $\Omega$-dependence in the Velocity Divergence Field} 

In this section, we extend the results obtained in \S 2 to the case where
the density parameter $\Omega \neq 1$. 
In general, an overall factor 
$f(\Omega) = d\,\log D(t)/d\,\log a(t)$ (see Eq.[\ref{fomega}] below), 
couples to the velocity divergence in the continuity equation 
for a non-flat space -what we shall call the {\em strong}
$\Omega$ dependence-, which scales in the
moments of the velocity divergence  
$\langle \theta^J \rangle \propto  f(\Omega)^J$. Therefore,
for the cumulants, one expects
\beq
T_J (\Omega) \approx {1 \over f(\Omega)^{J-2}}\,T_J (\Omega=1),
\label{strongom}
\eeq
with $f(\Omega) \approx \Omega^{0.6}$, provided one assumes a
vanishing cosmological constant $\Lambda = 0$ (see Peebles 1980). 
Nonetheless, there is also a {\em weak} 
dependence on the density parameter that modulates the dominant (strong)
dependence (see BCHJ95 and B94a). 
This dependence is induced by the
density field in a non-flat space and is obtained through the equation
of motion when the generalized coefficients of the density field 
$\nu_k (\Omega)$ replace those for flat-space in Eq.[\ref{munus}].  
However, as commented above (see \S 3), the fits to the $\nu_k (\Omega)$
coefficients can be derived either in the SC model approach or in the 
Lagrangian PT approach since they are formally equivalent (see Appendix  
\ref{ap:omegadep} for details).
   
From the fits to the SC model approach for the density field (see
\S\ref{omscapp}), we induce the following approximate $\Omega$-dependences
for the velocity divergence field whenever $2 \simgt \Omega \simgt 0.1$, 
\bea
{-T_3\, f(\Omega)} &=& 3\,\overline{\mu_2 (\Omega)} 
\approx \,{26\over 7} +
{12\over 7}\,
\left(\Omega^{-3/91}-1 \right) +\gamma  \nn \\
T_4 \,f(\Omega)^2 &=& 4\,\overline{\mu_3 (\Omega)} +
12\,\overline{\mu_2 (\Omega)}^2  \nn \\
&\approx&  4\,\left[{3022\over 441}\, 
+ {12\over 49}\,\left(\Omega^{-6/91}-1 \right)+
{169\over 42}\,\gamma+{7\over 12}\,\gamma^2 \right. \nn \\
&+& \left. {13\over 7}\,\gamma\,\left(\Omega^{-3/91}-1 \right) \right] . 
\label{tjg}
\eea
Alternatively, when we make use of the $\nu_k(\Omega)$ as they are derived 
from the Lagrangian PT approach (see Appendix \ref{ap:omegadep}) we get,
\bea
{-T_3 \,f(\Omega)} & \approx & {26\over 7} +
{12\over 7}\,\left(\Omega^{-2/63}-1 \right) +\gamma  \nn \\
T_4 \,f(\Omega)^2 & \approx & {12088\over 441}\, 
+ {192\over 49}\,\left(\Omega^{-4/63}-1 \right)+{80 \over 21}\,
\left(\Omega^{-2/35}-1 \right)  \nn \\ 
&-& {8\over 9}\,\left(\Omega^{-4/77}-1 \right) 
+ {338\over 21}\,\gamma+{7\over 3}\,\gamma^2  \nn \\
&+& 
{4\over 7}\,\left(22+13\,\gamma \right) \cdot \left(\Omega^{-2/63}-1 \right),
\label{tjlpt}
\eea
where the strong $\Omega$-dependence, $f (\Omega) = d \log D(\psi)/d \log a$ 
(where $a$ is the scale factor and $D(\psi)$ is the growing mode of the linear
density contrast), 
can be exactly integrated  and yields for $\Omega < 1$,

\bea
f(\Omega) &=& 3\, \sec h  ({\psi/2})\,
\left((2+\Omega)\,\psi -6 \,\sqrt{1 - \Omega} \right) \over
\Omega\,\left(9\,\sinh({\psi/2})+ \sinh({3\,\psi/2})
-6\,\psi\,\cosh({\psi/2}) \right)  \nn
\eea
while for $\Omega > 1$, it has the form,
\bea
f(\Omega) &=&  3\,\sec({\psi/2})\,
\left((2+\Omega)\,\psi -6 \,\sqrt{\Omega-1} \right) \over
\Omega\,\left(9\,\sin({\psi/2})+ \sin({3\,\psi/2})
-6\,\psi\,\cos({\psi/2}) \right)
\label{fomega}
\eea   
where,
$\psi = \cosh^{-1} ({2/\Omega}-1)$ .
However, the fit $f(\Omega) \approx \Omega^{3/5}$ is a good approximation
for open models as originally put forward by Peebles (1980). Other
more accurate fits are $f(\Omega) \approx \Omega^{13/22}$ for open models
and $f(\Omega) \approx \Omega^{13/23}$ for closed models within the range
$2 \simgt\Omega \simgt 1$.




\begin{figure}
\centering
\centerline{\epsfysize=8.truecm 
\epsfbox{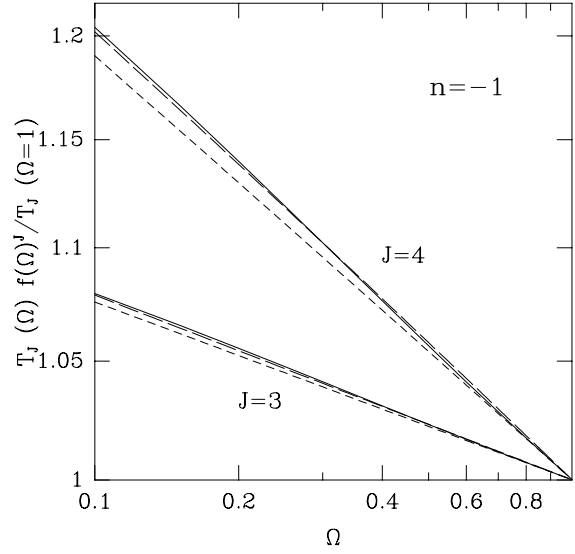}}
\caption[junk]{Same as Fig \ref{nftree_den} for the weak $\Omega$-dependence
of the velocity divergence (\ie excluding the overall 
$\sim f(\Omega)$ factor).}
\label{nftree_vel}
\end{figure}

In Fig \ref{nftree_vel} we plot the weak 
$\Omega$ dependence (\ie excluding the overall $\sim f(\Omega)$ factor)
for the hierarchical amplitudes at tree-level,
for a value of the spectral index $n=-1$.  
It shows two independent fits, the Lagrangian PT one 
(based on the spherically symmetric approximation to the perturbative 
expansion of the displacement field) 
and those within the SC model approach (same approximation but applied to the 
expansion of the density contrast), 
against the exact
behavior numerically integrated. It shows that the BCHJ95 fits to the
$g_k$ functions (see Eq.[\ref{gs}]) systematically 
underestimates the exact (weak) $\Omega$-dependence 
of the hierarchical amplitudes at tree-level, $T_3 f$
and $T_4 f^2$,
for open models 
(in the range $1 \simgt \Omega \simgt 0.1$) up to a 
$5 \%$ and $10 \%$, respectively. 
For both cases the fits found in the SC model approach are always less than
$1 \%$ away from to the exact behavior.

\begin{figure}
\centering
\centerline{\epsfysize=8.truecm 
\epsfbox{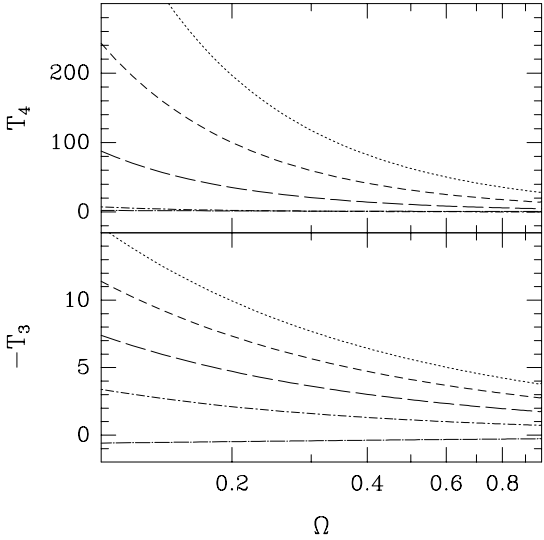}}
\caption[junk]{Values for the skewness and kurtosis of the velocity divergence
at tree-level
for open models. The smoothing effects strongly suppress the 
$\Omega$-dependence, which increases with the order of the
hierarchical amplitudes $T_J$. The dotted line displays the $n=-3$ (unsmoothed)
case while lower lines depict the smoothed cases for $n=-2,-1,0,1$, 
respectively,
from top to bottom. 
}  
\label{nflatvel}
\end{figure}

Fig \ref{nflatvel} displays the $\Omega$-dependence (combining
both the strong and weak dependences) in the skewness 
and kurtosis of the velocity divergence for different values of the
spectral index. There are two main features that arise in these 
dependences. 
First, although the relative variation of $T_J$ with $\Omega$ 
increases with the
smoothing effects (larger spectral index), the absolute variation decreases
with them, what makes it more plausible to probe $\Omega$ from lower 
values of the spectral index (which effectively means looking 
at small scales). Secondly,
the dependence with the density parameter increases with $J$, so that 
looking at higher-order moments of the velocity divergence in the
galaxy catalogues seems to be a promising test for determining $\Omega$.

We stress that the behavior of the closed models is just an
extrapolation of the one seen for the open models near $\Omega=1$, so that only
small changes arise in the range $10 \simgt \Omega \simgt 1$, as 
compared to the case of
the open models.

\subsection{ The $\Omega$-dependence for Non-Gaussian Initial Conditions}

For both cosmic fields (density or velocity)
the  $\Omega$-dependence with NGIC (for dimensional scaling,
$\langle \delta^J \rangle_c \propto {\langle \delta^2 \rangle}^{J/2}$) 
can be recovered by combining
in a straightforward way the results we have 
presented so far.

This can be done by noticing that the NGIC
results are given in terms of the Gaussian tree-level amplitudes
(see \S2.3 for the velocity and Paper II for the density).
Thus, to obtain the  $\Omega$-dependence one just have to
substitute the Gaussian tree-level amplitudes 
(\ie $S_J^G$ or $T_J^G$) with the corresponding  $\Omega$-dependent
amplitudes, given above in Eq.[\ref{sjg}] or  Eq.[\ref{tjg}]. 
For example the variance $\sigma^2$
of the density field for NGIC is:
 \bea
\sigma^2 &=& \sigma^2_l + s_{2,3}\,\sigma^3_l \, + \cdots \nn \\
s_{2,3}  &=& 
\left({13\over 21} + 
{6\over 21}\,\left( \Omega^{-3/91}-1 \right) + {\gamma\over{3}}\right)
\,B_3^0 
\eea
where $B_3^0 \equiv 
{\langle \delta_0^3 \rangle} /
{\langle \delta_0^2 \rangle}^{3/2}$ corresponds
to the amplitude of the (dimensional) skewness
of the initial density field.

\section{Discussion and Conclusions}
\label{discuss}

For Gaussian initial conditions (GIC), we 
found in Paper I, that the Spherical Collapse (SC) model gives
very accurate predictions for the hierarchical amplitudes $S_J$
in the perturbative regime, as compared
to the results derived by  
Scoccimarro \& Frieman (1996a, 1996b) for the exact PT 
in the diagrammatic approach (see Paper I \S 5.1).
We also compared the predictions for the higher-order
moments from the SC model to those measured in 
CDM and APM-like N-body simulations,
and they turned out to be in  
very good agreement in all cases up
to the scales where $\sigma_l \approx 1$, supporting our view that
the tidal effects only have a marginal contribution to the reduced
cumulants (see Paper I \S 5.2 \& \S 5.3). 
For Non-Gaussian initial conditions (NGIC) the SC
recovers all terms given by the exact PT 
for the variance, $\sigma$, skewness, $S_3$, and kurtosis, $S_4$,
 up to non-local integrals, involving initial
$n$-point functions that arise as a result of the coupling between the
asymmetric initial conditions with the tidal forces (see Paper II \S 3.4). 
The measured higher-order moments in the N-body simulations 
with NGIC (with dimensional, texture-like, scaling)
turned out to be in good agreement with predictions from the
SC model (see Paper II \S 3.3),
what gives further support to the domain of applicability of the
SC model in PT.

In this paper, we have extended the work presented in Papers I \& II
and have applied the SC model in Lagrangian space 
to study the predictions for the cumulants of the  velocity divergence field,
as well as the $\Omega$-dependence of the one-point cumulants 
in PT.
We provide with the results for the velocity field 
up to one loop (first correction to the leading term).      
However, the loop predictions turned out not to be as accurate as 
those for the density field, 
suggesting they are dominated by tidal effects. 
This can be explicitly 
seen in the variance, see (Eqs.[\ref{scvel}]-[\ref{ptvel}]) where the 
first non-linear
correction to the linear contribution is only accurate within an order 
of magnitude with respect to the exact result for the unsmoothed field.
The situation might be different for
the smoothed cumulants, as we have seen (in Papers I\&II) that 
for $n \simeq -1.5$ tidal effects seem to cancel out, at least in the
case of the density.
The SC loop predictions
for the {\em reduced} cumulants, are in better agreement with 
those obtained in the exact PT as comparison with the results
of Scocimarro \& Frimean (1996a) and Protogeros \& Scherrer (1997)
show. The latter suggests a cancellation
of the tidal contributions in line with the behavior found for the 
density field.  

As for the $\Omega$-dependence, we showed that the SC model is the exact 
solution at tree-level 
for the one-point cumulants for arbitrary density parameter, $\Omega$
(see also \S\ref{ap:scomega} for a demonstration in the context of B92). 
This property was explicitly used to 
derive the $\Omega$-dependence of the cumulants at tree-level 
for the density and velocity divergence fields. 
As an independent check to our results we worked out the 
{\em spherically symmetric} solution to Lagrangian PT, which was shown to be 
equivalent to our solution within the SC model in Lagrangian space as well
(see \S\ref{ap:omegadep}).
This way we were able to compare available fits to the $\Omega$-dependence
of the hierarchical amplitudes at tree-level
in the Lagrangian PT approach to our fits within the SC model approach. 
We found agreement within a few percent between these results as well as
with the exact numerically integrated solution 
(see Figs. \ref{nftree_den} and \ref{nftree_vel}
respectively). 

We also present, as a new result, 
the $\Omega$-dependence of the one-loop corrections 
in the SC approximation, for GIC and NGIC.
We show that both the variance and
the $S_J$ ratios increase as $\Omega$  decreases ($\Omega<1$).
 For the variance with GIC, this
trend is in line with fitting formulas and numerical
simulations (Peacock \& Dodds 1996) and reflects the
slowed-down growth of perturbations on large scales for high density
FRW universes. 
Although this effect is small (less than $10 \%$ for $\sigma<1$)
the SC results should be accurate for $n \simeq -1.5$, 
and could therefore be useful when compared to observations  
of the large scale structure
in the weakly non-linear regime.

\section*{Acknowledgments}

We want to thank Roman Scoccimarro, Francis Bernardeau
and Josh Frieman 
for carefully reading the 
manuscript and pointing out useful remarks. 
EG acknowledges support from CIRIT (Generalitat de
Catalunya) grant 1996BEAI300192. 
PF acknowledges a PhD grant supported by 
CSIC, DGICYT (Spain), projects PB93-0035 and PB96-0925.
This work has been
supported by CSIC, DGICYT (Spain), projects
PB93-0035, PB96-0925, and CIRIT, grant GR94-8001.

\section{References}

\def\refe {\par \hangindent=.7cm \hangafter=1 \noindent}
\def\aj { ApJ, }
\def\aa {A \& A, }
\def\prl {Phys.Rev.Lett.,}
\def\ajs{ ApJS, }
\def\mn { MNRAS, }
\def\mnl { MNRAS.Lett., }
\def\apl { ApJ.Lett., }

\refe Baugh, C.M., Gazta\~{n}aga, E., Efstathiou, G., 1995, \mn 274, 1049 
\refe Bernardeau, F., 1992, \aj 392, 1 (B92) 
\refe Bernardeau, F., 1994a, A\&A 291, 697 (B94a) 
\refe Bernardeau, F., 1994b, \aj 433, 1 (B94b)
\refe Bernardeau, F., 1994c, \aj 427, 51 
\refe Bouchet, F. R., Juszkiewicz, R., Colombi, S., 
1992 \apl 394, 5 
\refe Bouchet, F. R., Colombi, S., Hivon, E., Juszkiewicz, R.,
1995 \aa 296, 575 (BCHJ95)
\refe Catelan, P., Lucchin, F., Matarrese, S., Moscardini, L., 1995, \mn
276, 39 
\refe Chodorowski, M.J., Bouchet, F.R., 1996, \mn, 279, 557 
\refe Fosalba, P., Gazta\~{n}aga, E., 1998, \mn in press (Paper I, this issue)
\refe Fry, J.N., 1984, \aj 279, 499
\refe Fry, J.N., Scherrer, 1994, R.J., \aj 429, 1 
\refe Gazta\~naga, E. \&  Baugh, C.M., 1995, \mnl 273, 1
\refe Gazta\~naga, E. \& M\"{a}h\"{o}nen, P., 1996, \apl 462, 1
\refe Gazta\~naga, E., Fosalba, P., 1998, \mn in press (Paper II, this issue)
\refe Goroff, M.H., Grinstein, B., Rey, S.J., Wise, M.B., 1986, \aj 311, 6
\refe Juszkiewicz, R., Bouchet, F.R., Colombi, S., 1993 \apl 412, 9
\refe Lahav, O., Lilje, P.B., Primack, J.R., Rees, M.J., 1991, \mn 251, 128
\refe Martel, H., Freudling, W., 1991, \aj 371, 1
\refe Moutarde, F., Alimi, J.M., Bouchet, F.R., Pellat, R., Ramani, A., 
1991 \aj 382, 377
\refe Nusser, A., Colberg, J.M., 1998, \mn 294, 457
\refe Peacock, J.A., Dodds, S.J., 1996, \mn 280, 19. 
\refe Peebles, P.J.E., 1980, {\it The Large Scale Structure of the 
Universe:} Princeton University Press, Princeton
\refe Protogeros, Z.A.M., Scherrer, R.J., 1997, \mn 286, 223  
\refe Scoccimarro, R., Frieman J., 1996a, \ajs 105, 37 
\refe Scoccimarro, R., Frieman J., 1996b, \aj 473, 620 
\refe Silk J., Juszkiewicz, R.  1991, Nature, 353, 386 
\refe Zel'dovich, Ya.B., 1970, \aa 5, 84

\appendix 

\section{Gaussian Tree-Level with Arbitrary $\Omega$ }
\label{ap:scomega}

We want to show here, in the framework of Bernardeau (1992)
calculations, that 
the equations of motion
that govern the tree-level amplitudes for Gaussian initial conditions in a 
FRW universe with arbitrary density parameter, $\Omega$, 
are those of the SC dynamics.

Let us first introduce the vertex generating 
function which 
is defined as (see B92),
\bea
{\calG_{F}} (\tau)= \sum_{k=1}^{\infty}\,{\nu_k \over k!} \,
(-\tau)^k \nn
\eea
where $\nu_k = {\langle F^{(n)} \rangle^n_c}$, for a generic field $F$ and 
the superscript $n$ in the average denotes that
all tree-level contributions come through diagrams that have $n$
external lines (see B92 for details). 
According to B92, the equations of motion
for ${\calG_{\delta}}$ have the form,
\bea
\left[a\,{\partial \over \partial a}\, +
f(a)\tau\,{\partial \over \partial \tau}\right]\,{\calG_{\delta}} (a,\tau) = -
\left[1+{\calG_{\delta}} (a,\tau)\right]\,{\calG_{\Delta \Phi}} (a,\tau) ,  
\label{nonfsc1} \nn \\
\left[a\,{\partial \over{\partial a}}\,+
f(a)\tau\,{\partial \over{\partial \tau}} \right] 
\,{\calG_{\Delta \Phi}} (a,\tau) 
+ {1 \over{3}}\,{\calG^2_{\Delta \Phi}} (a,\tau) \, = \nn \\ 
\hspace{-1cm} - \left[ {2\nu-1 \over{\nu}} + 
{d\,\log \nu \over{d\,\log a}} \right] \,
{\calG_{\Delta \Phi}} (a,\tau) 
-{3 \over 2}\,\Omega\,{\calG_{\delta}} (a,\tau) \
\label{nonfsc2}   
\eea
being, $\Phi$ the peculiar velocity potential, 
$f(a) \equiv d\,\log D(a)/d\,\log a$, 
and $\nu \equiv d\,\log a/d\,\log t = H\,t$, where $t$ is the comoving time.
First, as we want to recover the equations of motion in terms of
the density contrast and the velocity divergence, we shall assume at
the tree-level, ${\calG_{\delta}} (a,\tau)= \delta$, as pointed out by B94 
although 
there it was restricted to flat-space. Replacing this last 
identity in 
the continuity equation [\ref{conteq}] leads to ${\calG_{\Delta \Phi}} =
\theta \equiv
\nabla \cdot \vv / f(a)\,H$ provided we introduce the
comoving time derivative in the way,
\bea
{d \over dt} \equiv H \,\left(a\,{\partial \over \partial a}\,+
f(a)\tau\,{\partial \over \partial \tau} \right). 
\eea
 Substituting this into (\ref{nonfsc2}), we get
\beq
{\ddot{\delta}}\,+2\,H \,{\dot{\delta}}\,-\,{4\over 3}\,{{\dot{\delta}}^2 
\over (1+\delta)}\,=\,{3\over 2}\,\Omega H^2 \delta (1+\delta) \,=\,
4\pi G \rhobar \delta \,(1+\delta) ,
\label{nonfsc3}
\eeq
where in the first equality we have made use of the equation of motion,
${\dot{H}}= -\left(1+{\Omega/2}\right)\,H^2$ 
valid for a $\Lambda =0 $ universe, while
the second equality follows from the Friedmann equation. 
Equation [\ref{nonfsc3}] is exactly the equation that
describes the evolution of density perturbations
in the SC model (see Paper I \S 4.1).

 On the other hand, since in the SC model in the perturbative regime
the evolved density field is given by
a local-density transformation of the kind,
\beq
\delta  =  f(\delta_l) =  
\,\sum_{n=1}^{\infty} {c_n \over n!}\, {[\delta_l]^n} ,
\eeq
from the above equivalence between the vertex
generating function and the density field from the SC model at tree-level,

\beq
\delta (\delta_l) = {\calG_{\delta}} (a,-\tau),
\label{gdelta}
\eeq
it follows that, $c_k (\Omega)= \nu_k (\Omega)$,
\ie the coefficients of the local-density transformation
are those of the SC model for a FRW universe with arbitrary 
$\Omega$.

\section{The $\Omega$-dependence in Lagrangian Perturbation Theory}
\label{ap:omegadep}

In this section we give a brief account of the Lagrangian PT which 
is based on an expansion of particle trajectories around the initial 
positions. We proceed and show the formal equivalence of the latter 
to the SC model in Lagrangian space. This result along with
the available fits to the Lagrangian PT found by Bouchet \etal (1995,
BCHJ95 hereafter) allows to extend their results for the skewness,
to the kurtosis of the smoothed density and velocity fields.

 Lagrangian PT contains the ZA as the first order of the
perturbative series. Higher-orders were considered 
by Moutarde \etal (1991), and were
later generalized by Bouchet \etal (1992). Our analysis here 
essentially follows the steps of BCHJ95 were the perturbative
analysis was extended up to third order.

 We start by introducing the density contrast and relating it to the 
displacement field, i.e., the perturbation around the original trajectory
in Lagrangian space,
\bea
\delta = {\rho \over \overline{\rho}}-1 = {1\over J}-1 \nn ,
\eea
where $J$ is the Jacobian of the transformation between real (Euler) and
Lagrangian space, $x(t,q) = q + \Psi (t,q)$,
\bea
J_{i j} &=& {\partial x_i \over \partial q^j} = \delta_{ij} + \Psi_{i,j}
\nn \\
J &=& \left| {\partial x \over \partial q} \right|  \nn ,
\eea
being $\Psi_i$ the component of the displacement field along the $i$ direction.

Expanding $J$ and, consequently, $\Psi_i$ in a perturbative series,
\bea
J &=& 1 + \epsilon J^{(1)} + \epsilon^2 J^{(2)} + \cdots     \nn \\ 
\Psi &=& \epsilon \Psi^{(1)} + \epsilon^2 \Psi^{(2)} + \cdots    
\eea
and rewriting $J$ in terms of the perturbative contributions from 
the displacement
field, we have,
\beq
J = 1 + \epsilon K^{(1)} + \epsilon^2 (K^{(2)} + L^{(2)}) + 
 \epsilon^3 \,(K^{(3)} + L^{(3)} + M^{(3)})+ \cdots 
\eeq
(see also Eq.[17] in Bernardeau 1994c) where the invariant scalars $K,L,M$ are,
\bea
K  &=& \nabla \cdot \Psi = \sum_{i} \Psi_{i,i} \nn \\
L &=& {1\over 2}\,\sum_{i,j} ( \Psi_{i,i}\,\Psi_{j,j} - \Psi_{i,j}\,\Psi_{j,i})
\nn \\
M &=& \det [\Psi_{i,j}]
\eea
form which the first perturbative orders are read off, 
\bea
K^{(m)} &=& \sum_{i} \Psi_{i,i}^{(m)} \nn \\ 
L^{(2)} &=& {1\over 2}\,\sum_{i \neq j} ( \Psi_{i,i}^{(1)}\,\Psi_{j,j}^{(1)} - 
\Psi_{i,j}^{(1)}\,\Psi_{j,i}^{(1)}) \nn \\
L^{(3)} &=& \sum_{i \neq j} ( \Psi_{i,i}^{(2)}\,\Psi_{j,j}^{(1)} - 
\Psi_{i,j}^{(2)}\,\Psi_{j,i}^{(1)}) \nn \\
M^{(3)} &=& \det [  \Psi_{i,j}^{(1)}] \nn .
\eea
The fluid equations of motion in Lagrangian space for a collisionless fluid
can be written down in the simple form (see BCHJ95 for details),
\beq
J(\tau,q) \,\nabla \ddot{x} = \beta(\tau)\,[J(\tau,q) -1],
\label{lagreqs}
\eeq

where $d\tau \propto a^{-2}\,dt$, and the dot denotes time derivatives with 
respect to $\tau$, while,
\beq
\beta(\tau) = {6 \over \tau^2 + k(\Omega)},
\eeq
with $k(\Omega)$ is a function of the geometry of the universe,
\bea
k(\Omega = 1) &=& 0 \nn \\
k(\Omega < 1) &=& -1 \nn \\
k(\Omega > 1) &=& +1 \nn .
\eea
The conformal time parameter $\tau=t^{-1/3}$ for a flat universe, and is
defined as
$\tau=\mid 1 - \Omega \mid^{-1/2}$, for non-flat universes.
Equation [\ref{lagreqs}] has a unique solution provided 
the fluid is irrotational
$\nabla\times\ddot x =0$, what we shall assume in the following. 

 Introducing the perturbative series in the equations 
of motion [\ref{lagreqs}],
one finds as generic solutions to the perturbed equations of motion for an
irrotational fluid,
\bea
K^{(1)} (\tau,q) = g_1 (\tau)\,K^{(1)} (\tau_i,q) \nn \\
K^{(2)} (\tau,q) = g_2 (\tau)\,K^{(2)} (\tau_i,q) \nn \\ 
K^{(3)} (\tau,q) = g_{3a}(\tau)\,M^{(3)} (\tau_i,q) + g_{3b} (\tau)\,
 L^{(3)} (\tau_i,q) . 
\eea
The first two orders are separable, unlike the
third one, where two independent growing modes have to be taken
into account. The above expression shall be taken as a definition of the
$g_k$ functions which encode the $\Omega$-dependence of the Jacobian $J$,
and thus that of the equations of motion in Lagrangian PT. 

We now impose the {\em spherical symmetry} 
in the growth of perturbations which, in terms of the displacement field 
reads,
\bea
& \Psi_{1,1} =  \Psi_{2,2} = \Psi_{3,3} = \Psi_{k,k} \nn \\
& \Psi_{i,j} = 0  \quad for \quad i \neq j.
\eea
Applying these simple properties on the perturbative series of the Jacobian
we find,

\bea
K^{(1)} (\tau,q) &=& 3\,\Psi_{k,k}^{(1)} (\tau,q) = 3\,g_1 (\tau)\,
\Psi_{k,k}^{(1)} (\tau_i,q), \nn \\
K^{(2)} (\tau,q) &=& 3\,g_2(\tau)\,{\Psi_{k,k}^{(1)}}^2 (\tau_i,q) \nn \\
L^{(2)} (\tau,q) &=& 3\,g^2_1(\tau)\,{\Psi_{k,k}^{(1)}}^2 (\tau_i,q) \nn \\
K^{(3)} (\tau,q) &=& (g_{3a} (\tau) +g_{3b} (\tau))\,
{\Psi_{k,k}^{(1)}}^3 (\tau_i,q) \nn \\
L^{(3)} (\tau,q) &=& 6\,g_2(\tau)\,g_1(\tau)\,{\Psi_{k,k}^{(1)}}^3 (\tau_i,q)
\nn \\
M^{(3)} (\tau,q) &=& g^3_1(\tau)\,{\Psi_{k,k}^{(1)}}^3 (\tau_i,q),
\eea
where we have made use of the fact that $K^{(2)}(\tau_i,q) =L^{(2)}(\tau_i,q)$
(see BCHJ95) which yields $\Psi_{k,k}^{(2)} (\tau_i,q) = 
{\Psi_{k,k}^{(1)}}^2 (\tau_i,q)$. Replacing these last expressions in the
perturbative orders of the Jacobian one gets,

\bea
J^{(1)} (\tau,q) &=& 3\,g_1 (\tau)\,\Psi_{k,k}^{(1)} (\tau_i,q) \nn \\
J^{(2)} (\tau,q) &=& 3\,[g^2_1(\tau) + g_2(\tau)] 
\,{\Psi_{k,k}^{(1)}}^2 (\tau_i,q) \nn \\
J^{(3)} (\tau,q) &=& \left[ g_{3a} (\tau) +6\,g_{3b} (\tau) \right.
\nn \\
&+& \left. 
6\,g_2(\tau)\,g_1(\tau)+g^3_1(\tau)) \right]\, {\Psi_{k,k}^{(1)}}^3 (\tau_i,q).
\label{jpert}
\eea

On the other hand, the unsmoothed transformation 
of the evolved density contrast in the SC model reads,
\bea
\delta &=& \delta_l + {\nu_2 \over 2}\,\delta_l^2+
{\nu_3 \over 3!}\,\delta_l^3 + \cdots \nn 
\eea
while the analog expansion in terms of the Jacobian gives,
\bea
\delta &=& -\epsilon \,J^{(1)} + \epsilon^2\,({J^{(1)}}^2 - J^{(2)}) \nn \\
&+& 
\epsilon^3\,(-{J^{(1)}}^3 + 2\,J^{(1)}\,J^{(2)} - J^{(3)}) + \cdots \nn
\eea
from which we see that the linear term is just $\delta_l = -\epsilon\,J^{(1)}$,
and all perturbative orders can be expressed as powers of the linear term,
since $J^{(n)} \propto [J^{(1)}]^{n}$, in the same way
$\delta^{(n)} \propto [\delta_l]^{n}$ in the SC model.  

Identifying in the same way the higher-orders one by one (in powers of
$\epsilon$) and replacing the expressions for the $J^{(i)}$ according
to Eq.[\ref{jpert}], we arrive at,

\bea
\nu_2 &=& {2 \over 3}\,\left(2 - {g_2 \over g^2_1} \right) \nn \\
\nu_3 &=& {2 \over 9}\,\left(10 - 12\,{g_2 \over g^2_1} + 
{g_{3a} \over g^3_1} + 6\,{g_{3b} \over g^3_1} \right).
\eea
Making use of these coefficients we immediately obtain the tree-level
amplitudes for the skewness and kurtosis of the density field in terms of
the $g_k$ functions which, in turn, incorporate the $\Omega$-dependence,

\bea
S_3 &=& 3\,\nu_2 = 4 - 2\,{g_2 \over g^2_1} \nn \\
S_4 &=& 4\,\nu_3 + 12\,\nu_2^2 \nn \\
    &=& {8 \over 9}\,\left(34 - 36\,{g_2 \over g^2_1} 
+ 6\,\left({g_2 \over g^2_1}\right)^2 + {g_{3a} \over g^3_1} + 
6\,{g_{3b} \over g^3_1} \right) .
\eea
The expression for $S_3$ was already derived by BCHJ95 so 
the above expressions
extend their results to the kurtosis in a simple fashion. 
An immediate extension of these results to the {\em smoothed} fields is
achieved by applying the expressions for the smoothed
coefficients of the local non-linear transformation 
that describes the spherical collapse $\overline{\nu_k}$ 
(see \S 4.4 in Paper I),
\bea
S_3 &=& 3\,\overline{\nu_2} = 4 + \gamma -2\,{g_2 \over g^2_1} \nn \\
S_4 &=& 4\,\overline{\nu_3} + 12\,\overline{\nu_2}^2 \nn \\
&=& {1\over 9}\,\left[272 + 150\,\gamma + 21\,\gamma^2 -12\,(24 +7\,\gamma)
\,{g_2 \over g^2_1} \right. \nn \\
&+& \left. 48\,\left({g_2 \over g^2_1}\right)^2 + 
8\,{g_{3a} \over g^3_1} + 48\,{g_{3b} \over g^3_1} \right] ,
\eea
where the ZA (or the SSZA in our spherically symmetric approach), 
which corresponds to the
first perturbative order in the Lagrangian approach, is recovered by 
setting $g_k = 0$, i.e., the $\Omega$-independent term.
  
Though the exact analytic expressions for $g_2/g_1^2$ are known (see BCHJ95),
no exact results are available for $g_{3a}/g_1^3$ and $g_{3b}/g_1^3$, so we
shall focus on the numerical fits for the cosmologically favored 
parameter space for $\Omega$ and work out the predictions. According to
BCHJ95, near $\Omega = 1$ good fits are provided by the factors given by 
Eq.[\ref{gs}].
Replacing these expressions for the $\Omega$-dependence 
in the skewness and kurtosis, we finally
get the fits to the smoothed $S_J$ amplitudes given by Eq.[\ref{sjlpt}].

 
Similarly, we can derive the corresponding fits to the 
{\em weak} $\Omega$-dependence
in Lagrangian PT                           
for the skewness and kurtosis
of the velocity divergence just by recalling the relationship between the 
unsmoothed coefficients of the local transformations for the density and 
velocity field provided by Eq.[\ref{munus}], and the expressions for the
smoothed coefficients in terms of the unsmoothed ones provided by
Eq.[\ref{musm}].

The smoothed $T_3$ and $T_4$ then read,
\bea
-T_3 (\Omega)\,f(\Omega) &=& 2+\gamma-4\,{g_2\over g_1^2} \nn \\
T_4 (\Omega)\,f(\Omega)^2 &=& {1\over 3}\,\left[24+26\,\gamma+7\,\gamma^2
-4\,\left(22+13\gamma \right)\,{g_2\over g_1^2}\, \right. \nn \\
&+& \left. 64\,\left({g_2\over g_1^2}\right)^2+
8\,{g_{3a}\over g_1^3}+48\,{g_{3b}\over g_1^3}\right] ,
\eea
where the overall factor $f(\Omega)$ gives the strong dependence on the density
parameter given by Eq.[\ref{fomega}].
Introducing the  
fits given by BCHJ95 for the density field in the above given expressions, 
we obtain 
their approximated $\Omega$-dependences given by Eq.[\ref{tjlpt}].


\end{document}